\documentclass[preprint,showpacs,preprintnumbers,amsmath,amssymb,nofootinbib]{revtex4}
\usepackage{bbm}
\usepackage{amsfonts}
\usepackage{booktabs}
\usepackage{mathrsfs}
\usepackage{epsfig}
\usepackage{graphicx}
\usepackage{dcolumn}
\usepackage{bm}
\usepackage{amsmath}
\usepackage{slashed}

\let\jnfont=\rm
\def\NPB#1,{{\jnfont Nucl.\ Phys.\ B }{\bf #1},}
\def\PLB#1,{{\jnfont Phys.\ Lett.\ B }{\bf #1},}
\def\EPJC#1,{{\jnfont Eur.\ Phys.\ Jour.\ C }{\bf #1},}
\def\PRD#1,{{\jnfont Phys.\ Rev.\ D }{\bf #1},}
\def\PRL#1,{{\jnfont Phys.\ Rev.\ Lett.\ }{\bf #1},}
\def\MPLA#1,{{\jnfont Mod.\ Phys.\ Lett.\ A }{\bf #1},}
\def\JPG#1,{{\jnfont J.\ Phys.\ G}{\bf #1},}
\def\CTP#1,{{\jnfont Commun.\ Theor.\ Phys.\ }{\bf #1},}
\def\ZPC#1,{{\jnfont Z.\ Phys.\ C }{\bf #1},}
\def\JHEP#1,{{\jnfont JHEP \ }{\bf #1},}
\def\Rv{\not{\hbox{\kern-1pt $R$}}}
\def\p{\not{\hbox{\kern-3pt $p$}}}

\begin{document}
\preprint{\parbox{1.2in}{\noindent arXiv:}}

\title{Two-Higgs-doublet model with a color-triplet scalar: a joint explanation for
top quark forward-backward asymmetry and Higgs decay to diphoton}

\author{Chengcheng Han$^{1,2}$, Ning Liu$^1$, Lei Wu$^{2}$, Jin Min Yang$^2$, Yang Zhang$^1$
        \\~ \vspace*{-0.3cm} }
\affiliation{$^1$ Physics Department, Henan Normal University, Xinxiang 453007, China\\
$^2$ State Key Laboratory of Theoretical Physics,
      Institute of Theoretical Physics, Academia Sinica, Beijing 100190,
      China \vspace*{1.5cm}}

\begin{abstract}
The excess of top quark forward-backward asymmetry ($A^t_{FB}$)
reported by the Tevatron and the enhancement of the Higgs decay to
diphoton observed by the LHC may point to a same origin of new
physics. In this note we examined such anomalies in the
two-Higgs-doublet model with a color-triplet scalar. We found that
under current experimental constraints this model can simultaneously
explain both anomalies at $1\sigma$ level. Also, we examined the
Higgs decay $h\to Z\gamma$ and displayed its correlation with $h\to
\gamma\gamma$. We found that unlike other models, this model
predicts a special correlation between  $h\to Z\gamma$ and $h\to
\gamma\gamma$, i.e., the $Z\gamma$ rate is highly suppressed while
the $\gamma\gamma$ rate is enhanced. This behavior may help to
distinguish this model in the future high luminosity run of the LHC.

\end{abstract}

\maketitle

\section{INTRODUCTION}
Although the Standard Model (SM) agrees quite well with collider
experiments, new physics is speculated to appear at TeV scale. If
new physics beyond the SM indeed exist, it may readily affect the
Higgs sector and the top quark sector. The reason is that the Higgs
sector is responsible for electroweak symmetry breaking while the
top quark is the heaviest fermion and sensitive to electroweak
symmetry breaking. Actually, the top quark and Higgs boson entangled
with each other, e.g., they couple 'strongly', the top quark loop
dominates the Higgs production $gg\to h$ and also sizably impact the
clean decay $h\to\gamma\gamma$. So, the new physics (if really
exist) may be simultaneously present in both sectors.

Luckily, we have seen anomalies from both the LHC Higgs data and the Tevatron
top quark data. From the LHC Higgs data \cite{atlas-higgs-7,cms-higgs-7} we
see an enhancement in the diphoton channel while from the Tevatron top quark data
\cite{tev-afb} we see an excess for the top forward-backward asymmetry. Let's
take a look at these anomalies:
\begin{itemize}
\item[(i)] For the Higgs boson,
recently the ATLAS \cite{atlas-higgs-7} and CMS \cite{cms-higgs-7}
collaborations independently reported observation of a Higgs-like
resonance around 125 GeV. Their observations are also supported by
the Higgs search at the Tevatron \cite{tev-higgs}.
Although the property of this Higgs-like boson is in rough agreement
with the SM Higgs prediction, an excess at 2$\sigma$ level was
observed in the diphoton channel \cite{atlas-rr,cms-rr} and an enhancement in
the $Vb\bar{b}$ channel was also reported by the Tevatron \cite{tev-vbb}.

\item[(ii)] For the top quark, although most measurements at the
Tevatron and LHC are well consistent with the SM predictions,
an excess for the top quark forward-backward asymmetry ($A^t_{FB}$)
was observed by both CDF and D0 collaborations.
For $A^t_{FB}$ in the inclusive $t\bar{t}$ production and in the high
$t\bar{t}$ invariant mass region ($m_{t\bar{t}}>450$ GeV), the deviations
from the SM predictions \cite{afb-sm} are about $1.5\sigma$ and $2.4\sigma$
respectively \cite{tev-afb}.
\end{itemize}

Although these anomalies are quite mild and may go away in the future,
they stimulated various new physics explanations.
The enhancement of the Higgs diphoton signal can be explained in
the popular low energy supersymmetry (SUSY) \cite{diphoton-SUSY}
and other miscellaneous models (say with new scalars,
new fermions or new vector bosons)
\cite{Carena,Djouadi,new-scalar,new-fermion,new-vector,new-tensor}
(note that the Higgs diphoton rate cannot be enhanced in
some well-known new physics models like little Higgs theory and universal
extra dimensions \cite{diphoton-lht}).
For the anomaly of $A^t_{FB}$, the popular new physics models like SUSY
cannot explain it while other strange models are tried to provide
an explanation \cite{afb-review,afb-axigluon,afb-scalar-vector,afb-eft}.

It is noteworthy that for  the anomaly of $A^t_{FB}$ a diquark model with
a color anti-triplet scalar
($\phi$) can provide a nice explanation \cite{afb-diquark} although
it will inevitably lead to a color triplet resonance
in the $\bar{t}(t)+j$ system of the $t\bar{t}$ + jets signature and
thus highly constrained by the current data of top pair or
single top production at the LHC \cite{cms-tt,atlas-t}.
The intriguing point of this model is that the new predicted
charged and colored scalar can also affect the
processes $gg\to h$ and $h\to \gamma\gamma$ through
the Higgs portal operator $\phi^{\dagger} \phi
H^{\dagger}H$ ($H$ is the SM Higgs doublet) and may
enhance the Higgs diphoton rate at the LHC.
Actually, the effects of the general charged and colored scalars
in the Higgs productions and decays have been studied in
\cite{Donser}, where it is found that the scalars transforming as
$(8,2,1/2)$ and $(\bar{6},3,-1/3)$ under the SM gauge group can
improve the agreement with the Higgs data.
Meanwhile, if the Higgs sector of the SM is also extended,
such as in SUSY, a light color-triplet stop will be
feasible for improving the fit of Higgs data \cite{susy}.
So the extension of the Higgs sector may also play an important role.

In this note, we focus on the two-Higgs-doublet model with a
new color anti-triplet scalar and examine the top quark
forward-backward asymmetry and the Higgs diphoton decay.
Under all available experimental constraints, we scan the
parameter space to figure out if this model can provide a joint explanation
for the anomalies of $A^t_{FB}$ and Higgs decay to diphoton.
Also, we will study the Higgs decay $h\to Z\gamma$ and
display its correlation with $h\to \gamma\gamma$.

This paper is organized as follows. In Sec.
II, we briefly outline the relevant features of the diquark model.
In Sec. III we examine the experimental constraints on the
parameter space of the diquark model and then in the allowed parameter
space we calculate $h \to \gamma\gamma$, $h\to Z \gamma$ and $A^t_{FB}$.
Finally, we draw our conclusion in Sec. IV.

\section{A brief description of the model }
Motivated by the top quark forward-backward asymmetry, we consider a
diquark model, in which the new scalar $\phi$ transforms as
($\bar{3}$,1,\-4/3) under the gauge group $SU(3)_C\times SU(2)_{L}
\times U(1)_{Y}$ of the SM. Such a scalar can appear as a
pseudo-Goldstone boson after the condensation of the
fourth family fermions \cite{4gen}
or the condensation of the stop in SUSY with large
$A_t$ term \cite{stopstop}. Compared with other diquarks with
exotic quantum numbers, such as the color-sextet scalar, the
color-triplet diquark is more favored by the measurement of $t\bar{t}$
cross section because its interference with the SM
processes is destructive \cite{afb-diquark}. After the electroweak
symmetry breaking, the relevant Lagrangian of the color anti-triplet
is given by \cite{afb-diquark}
\begin{eqnarray} \mathcal {L}\supset
f_{ij}
\bar{u}_{i\alpha}P_{L}u^{c}_{j\beta}\epsilon^{\alpha\beta\gamma}
\phi^{\dagger}_{\gamma} + \hbox{h.c.}\;,  \label{phi}
\end{eqnarray}
where $f_{ij}=-f_{ji}$ with the indices $i$ and $j$ denoting the quark
flavors, $\epsilon^{\alpha\beta\gamma}$ is the antisymmetric tensor for
the color indices $\alpha,\beta,\gamma$, and $t^{c}$ is defined as
$t^{c}=C\bar{t}^{T}$ with $C$ being the charge conjugate operator. In
this model, since the main contribution to the process
$u\bar{u}\rightarrow t\bar{t}$ arises from flavor changing
interaction between up quark and top quark, for simplicity, we can
set the couplings $f_{cu,ct}=0$ to escape the constraints from low
energy flavor physics like $D^{0}-\bar{D^{0}}$ mixing \cite{afb-diquark}.
In this paper, we only use the low energy effective interaction of
$\phi$ and will not discuss the UV completion of this model.
Therefore, we treat the coupling $f_{ut}$ as a free parameter.

Another feature of this diquark model is that the color
triplet scalar $\phi$ can couple to the Higgs boson through the
Higgs portal operator $\phi^{\dagger}\phi H^{\dagger}H$.
In this work, considering the current Higgs data, we study the
diquark in the framework of two-Higgs-doublet
model, where the interaction between diquark and Higgs boson
is given by \cite{twosig}
\begin{eqnarray}
V
=&m_\phi^2\phi^{\dagger}\phi+\kappa^2|\phi^{\dagger}\phi|^2
+\lambda^u\phi^{\dagger}\phi|H_u|^2+\lambda^d\phi^{\dagger}\phi|H_d|^2
+\left ( \lambda_M\phi^{\dagger}\phi H_d\cdot H_u+h.c.\right)
\end{eqnarray}
with $H_u$ and $H_d$ denoting the two Higgs doublets.
After the electroweak symmetry breaking, the two Higgs doublets can
be written as
\begin{equation}
\label{conventions}
  H_d = \begin{pmatrix}
         v_d + (h_d + i a_d)/\sqrt{2} \\ H_d^-
        \end{pmatrix},\hspace{2mm}
 H_u = \begin{pmatrix}
         H_u^+ \\ v_u + (h_u + i a_u )/\sqrt{2}
        \end{pmatrix}, \hspace{2mm}
\end{equation}
where $v_u$ and $v_d$ are the vacuum expectation values.
Then we rotate these Higgs fields from the interaction eigenstates
to the mass eigenstates:
\begin{eqnarray}
\left( \begin{array}{c}   H \\ h \end{array} \right)
&=& \left( \begin{array}{cc} \cos \alpha & \sin \alpha \\
- \sin \alpha & \cos \alpha \end{array} \right) \ \left(
\begin{array}{c}   h_d \\ h_u \end{array} \right)
\end{eqnarray}
\begin{eqnarray}
\left( \begin{array}{c}   G^\pm \\ H^\pm \end{array} \right)
&=& \left( \begin{array}{cc} \cos \beta & \sin \beta \\
- \sin \beta & \cos \beta \end{array} \right) \ \left(
\begin{array}{c}   H_d^\pm \\ H_u^\pm \end{array} \right)
\end{eqnarray}
\begin{eqnarray}
\left( \begin{array}{c}   G^0 \\ A \end{array} \right)
&=& \left( \begin{array}{cc} \cos \beta & \sin \beta \\
- \sin \beta & \cos \beta \end{array} \right) \ \left(
\begin{array}{c}   a_d \\ a_u \end{array} \right),
\end{eqnarray}
where $\tan\beta$ is defined as the ratio $v_u / v_d$ and $\alpha$
is the mixing angle between the two CP-even Higgs bosons. For
simplicity, we assume $\lambda^u$=$\lambda^d$=$\lambda$
and then obtain the interactions between the
diquark and the light CP-even Higgs:
\begin{eqnarray}
\lambda\phi^{\dagger}\phi|H_u|^2+\lambda\phi^{\dagger}\phi|H_d|^2& =&
\lambda\phi^{\dagger}\phi\left[H^+_dH^-_d+v^2_d+\frac{1}{2}(h^2_d+a^2_d)+
H^+_uH^-_u  \right.        \nonumber  \\
& &\left. +v^2_u+\sqrt{2}(v_uh_u+v_dh_d) +\frac{1}{2}(h^2_u+a^2_u)\right] \nonumber \\
&\supseteq& \sqrt{2}\lambda v
\mathrm{sin(\beta-\alpha)}\phi^{\dagger}\phi h+\lambda
v^2\phi^{\dagger}\phi \\
\lambda_M\phi^{\dagger}\phi H_d\cdot H_u+h.c.
&=&
\lambda_M\phi^{\dagger}\phi(2v_uv_d+h_uh_d-a_da_u+\sqrt{2}v_uh_d+\sqrt{2}v_dh_u \nonumber \\
& & -H_u^+H_d^--H_d^+H_u^-) \nonumber  \\
&\supseteq&
\sqrt{2}\lambda_M\mathrm{cos(\beta+\alpha)}v\phi^{\dagger}\phi
h+\lambda_M v^2 \mathrm{sin(2\beta)}\phi^{\dagger}\phi
\end{eqnarray}
Here we take the light CP-even Higgs boson as a SM-like Higgs boson.
From Eqs.(7) and (8), we obtain the coupling of $h\phi^{\dagger}\phi$:
\begin{eqnarray}
\mathcal {L}_{h\phi^{\dagger}\phi} &=& -\sqrt{2}v[\lambda
\mathrm{sin(\beta-\alpha)}+\lambda_M\mathrm{cos(\beta+\alpha)}]h\phi^{\dagger}\phi.
\end{eqnarray}
Then in term of an effective coupling $g_{h\phi\phi}$ and the diquark mass
$m_s$, we can parameterize the above interaction as
\begin{eqnarray}
\mathcal {L}_{h\phi^{\dagger}\phi} &=& -\sqrt{2}[\lambda
\mathrm{sin(\beta-\alpha)}+\lambda_M\mathrm{cos(\beta+\alpha)}]vh\phi^\dagger
\phi \equiv - g_{h\phi\phi} {\frac{2 m_s^2 }{v} } h \phi^\dagger \phi
\end{eqnarray}
where $m_s^2=m_{\phi}^2+\lambda v^2+\lambda_M v^2
\mathrm{\sin2\beta}$. For the reduced couplings of Higgs boson with
the vector bosons $c_{hVV}(V=Z,W)$ and fermions $c_{hf\bar{f}}$,
they are same as in the general THDM:
\begin{equation}
 c_{hVV} = \sin(\beta - \alpha)\,, \quad c_{hb\bar{b}} =
-{\sin \alpha \over \cos \beta}\,, \quad c_{ht\bar{t}} = {\cos
\alpha \over \sin \beta}\,.
\end{equation}
In order to avoid the constraints from flavor physics and the LHC
search for the non-standard Higgs bosons, we carry out the
calculations in the decoupling limit $m_A \gg m_Z$. Therefore, the
contribution of those non-standard Higgs bosons to the low energy
observables is decoupled in our study.

\section{NUMERICAL RESULTS AND DISCUSSIONS}
Due to the contributions of the new charged scalar, the decay
widths of $h\to gg$, $h\to \gamma\gamma$ and $h \to Z\gamma$ in the
SM will be modified. In our case, the
partial width of $h \to \gamma \gamma$ can be expressed as \cite{Carena}
\begin{equation}
\Gamma_{h\rightarrow\gamma\gamma}=\frac{G_\mu\alpha^2m_h^3}{2\sqrt{2}\pi^3}|
A( {\rm W-loop})+A({\rm top-loop})
+A({\rm diquark-loop})|^2
\end{equation}
where the amplitude functions are given by
\begin{equation}
\begin{aligned}
&A({\rm w-loop})= -\frac{7}{8}c_{hVV}A_v(\tau_W),\\
&A({\rm top-loop})=\frac{2}{9}c_{ht\bar{t}}A_f(\tau_{t}),\\
&A({\rm diquark-loop})= \frac{N(r_s)}{24}Q_s^2g_{h\phi\phi} A_s(\tau_{\phi})
\end{aligned}
\end{equation}
with $\tau_i=m_h^2/4m_i^2$ ($i=W,t,\phi$), $N(r_s)$ being the
dimension of the color representation of the new scalar and $Q_s$
being its electric charge. For our model, $N(r_s)=3$ and $Q_s=-4/3$. The
one-loop form factors are given by \cite{statusreport}
\begin{eqnarray}
    A_s(\tau) &=& \frac{3}{\tau^2}[f(\tau)-\tau],\\
    A_f(\tau) &=& \frac{3}{2\tau^2}[(\tau -1)f(\tau)+\tau],\\
    A_v(\tau) &=& \frac{1}{7\tau^2}[3(2\tau-1)f(\tau)+3\tau+2\tau^2],\\
    f(\tau)&=& \left\{
                   \begin{array}{ll}
                     \mathrm{arcsin}^2\sqrt{\tau}, & \hbox{$\tau \leq 1$;} \\
                     -\frac{1}{4}[\mathrm{log}\frac{1+\sqrt{1-\tau^{-1}}}{1-\sqrt{1-\tau^{-1}}}-i\pi]^2, & \hbox{$ \tau > 1 $.}
                   \end{array}
                 \right.
    \label{Afun}
\end{eqnarray}
The partial width of $h \to gg$ is given by \cite{Carena}
\begin{equation}
\Gamma_{h\to
gg}=\frac{G_\mu\alpha_s^2m_h^3}{36\sqrt{2}\pi^3}|A({\rm top-loop})
+A({\rm diquark-loop}|^2
\end{equation}
where the amplitude functions are
\begin{eqnarray}
&&A({\rm top-loop})=c_{ht\bar{t}}A_f(\tau_{t}),   \\
&&A({\rm diquark-loop})=\frac{1}{2}C(r_s)g_{h\phi\phi} A_s(\tau_{\phi})
\end{eqnarray}
with $C(r_s)=1/2$ being the quadratic Casimir of the color
representation of the diquark. At leading order, the cross section
of $gg\to h$ is directly related to the gluonic decay width
$\Gamma(h\to gg)$:
\begin{equation}
\hat{\sigma}_{gg\to h} = \frac{\pi^{2}}{8m_h}\Gamma_{h\to
gg}\delta(\hat{s}-m^{2}_{h}) .
\end{equation}
From Eqs.(12) and (14) we see that there are two ways to
enhance the diphoton rate. One way is that the contribution of
diquark is constructive to the $W$-loop, which needs a
negative $g_{h\phi\phi}$. However, in this case, the main
production process $gg \to h$ will be highly suppressed
and thus the Higgs data, especially for $h\to VV^*$ \cite{Donser}
can hardly be accommodated. The other way is that the contribution of
diquark is destructive to the $W$-loop. Then we need
a large positive $g_{h\phi\phi}$. This will also enhance the production
rate of $gg\to h$. We found this case to be consistent with the Higgs
data by enhancing the decay width of $h\to b\bar{b}$,
which is also needed to explain the observation of $h\to
b\bar{b}$ at Tevatron \cite{tev-vbb}.

Note that, due to the gauge symmetry,
$h\to \gamma\gamma$ and $h\to Z\gamma$ have a strong correlation in
the SM. Therefore, the precise measurement of $Z\gamma$ can help
to understand the diphoton anomaly.
The partial width of $\Gamma_{h\rightarrow Z\gamma}$ is given by \cite{Carena}
\small
\begin{eqnarray}
\Gamma_{h\to Z\gamma} = \frac{G_F^2 m_W^2 \alpha m_h^3}{64\pi^4}\left( 1-\frac{m_Z^2}{m_h^2} \right)^3
\left|A({\rm W-loop})+A({\rm top-loop})+A({\rm diquark-loop}) \right|^{2}
\label{hzgamma}
\end{eqnarray}
\normalsize
where the amplitude functions are
\begin{eqnarray}
&&A({\rm W-loop})=\cos\theta_{W}c_{hVV}\mathcal C_{v}(\tau_W, y^{-1}_{W}),   \\
&&A({\rm top-loop})={\frac{2-(16/3)\sin^{2}\theta_{W}}{\cos\theta_{W}}c_{ht\bar{t}}
\mathcal C_{f}(\tau_t,y^{-1}_{t})}, \\
&& A({\rm diquark-loop})=-\sin\theta_Wg_{Z\phi\phi}N(r_s)g_{h\phi\phi}\mathcal C_{s}(\tau_s,y^{-1}_{s}),
\end{eqnarray}
with $y_i=m_Z^2/4m_i^2$ for $i=W,t,\phi$ and
$g_{Z\phi\phi}=2(T_\phi^3-Q_s\sin^2\theta_W)/\sin2\theta_W$. The
loop functions are given by
\begin{eqnarray}
&& \mathcal C_{s}(x,y)=I_{1}(x,y),   \nonumber\\
&& \mathcal C_{v}(x,y)=4(3-\tan^{2}\theta_{W})I_{2}(x,y)+\left((1+2x^{-1})\tan^{2}\theta_{W}-(5+2x^{-1})\right)I_{1}(x,y),\nonumber\\
&& \mathcal C_{f}(x,y)=I_{1}(x,y)-I_{2}(x,y),
\end{eqnarray}
where
\begin{eqnarray}
&&I_{1}(x,y)=\frac{x y}{2(x-y)}+\frac{x^{2} y^{2}}{2(x-y)^{2}}\left(f(x^{-1})-f(y^{-1})\right)+\frac{x^{2}y}{(x-y)^{2}}\left(g(x^{-1})-g(y^{-1})\right), \nonumber\\
&&I_{2}(x,y)=-\frac{x y}{2(x-y)}\left(f(x^{-1})-f(y^{-1})\right),  \nonumber\\
&&g(x)=\sqrt{x^{-1}-1}\arcsin\sqrt{x}.\;\;\;\;
\end{eqnarray}
In our study we consider the
following experimental results about the top quark and Higgs boson
from the LHC and Tevatron:
\begin{itemize}
\item[(i)] The cross section of $t\bar{t}$ production:
The CDF and CMS collaborations have measured the total $t\bar{t}$
cross sections respectively \cite{tev-tt-exp,cms-tt}, which are in
good agreement with the corresponding values predicted by the SM
\cite{sm-tt},
    \begin{eqnarray}
    && \sigma^{t\bar{t},Tev}_{exp} = 7.50 \pm 0.31 \pm 0.34 ~{\rm
    pb},~~~~\sigma^{t\bar{t},Tev}_{th} = 7.15^{+0.21+0.30}_{-0.20-0.25} ~{\rm pb} ;\nonumber\\
    && \sigma^{t\bar{t},LHC}_{exp} = 161.9\pm2.5 \pm3.6 ~{\rm pb},~~~~
    \sigma^{t\bar{t},LHC}_{th} = 162.4^{+6.7+7.3}_{-6.9-6.8} ~{\rm pb} .
    \end{eqnarray}
In our calculations, we require the theoretical prediction (the SM
value plus new physics effects) for the $t\bar{t}$ total cross
section to agree with the experimental data at $2\sigma$ level.
\item[(ii)]
Top+jet resonance in $t\bar{t}$+jets events: In our model, the top
quark can be produced in association with a diquark through new top
flavor violating interactions. This will lead to a resonance in the
top plus jet system of $t\bar{t}$+jets events when the diquark decays to
$\bar{t}q$.
  \begin{itemize}
  \item Tevatron:
The CDF Collaboration has recently searched for a $t ({\rm
or}~\bar{t})+$jet resonance in $t\bar{t}$+jet events and set an
upper limit of $0.61 \sim 0.02$ pb for $m_X=200 \sim 800$ GeV
\cite{tev-tj}.
 \item LHC: The similar search has been also carried out by the ATLAS collaboration
at the LHC \cite{atlas-tj}. The measurement is consistent with the
SM prediction and a color triplet scalar with mass below 430 GeV is
excluded at 95\% confidence level, assuming unit right-handed
coupling. However, this bound can be released when the coupling
become small.
  \end{itemize}

\item[(iii)] The charge asymmetry in $t\bar{t}$ production at the LHC:
The charge asymmetry in
$t\bar{t}$ production has been measured by the CMS \cite{cms-ac} and
ATLAS collaborations\cite{atlas-ac} in single lepton channel and is
consistent with the SM prediction \cite{sm-ac},
    \begin{eqnarray}
    && \sigma^{t\bar{t},CMS}_{exp} = 0.004 \pm 0.010 \pm 0.012;\nonumber\\
    && \sigma^{t\bar{t},ATLAS}_{exp} = -0.024 \pm 0.016 \pm 0.023;\nonumber\\
    && \sigma^{t\bar{t},LHC}_{th} = 0.0115 \pm 0.0006
    \end{eqnarray}
We also require the new physics contribution plus the SM value to
agree with the data at $2\sigma$ level.

\item[(iv)] The Higgs search data at the Tevatron and LHC:
For different decay modes of Higgs boson,
we define the following ratios of signal rate relative to the
SM prediction:
\begin{eqnarray}
&& R_{VV^*}^{LHC} \equiv
\frac{\sigma(ggh+VBF)}{\sigma^{SM}(ggh+VBF)}
\times \frac{Br(h\rightarrow VV^*)}{Br_{SM}(h\rightarrow VV^*)} \nonumber \\
&& R_{\gamma\gamma}^{LHC} \equiv
\frac{\sigma(ggh+VBF)}{\sigma^{SM}(ggh+VBF)} \times
\frac{Br(h\rightarrow \gamma\gamma)}{Br_{SM}(h\rightarrow
\gamma\gamma)} \nonumber \\
&& R_{Vbb}^{Tev} \equiv \frac{\sigma(Vh)}{\sigma^{SM}(Vh)} \times
\frac{Br(h\rightarrow b \bar{b})}{Br_{SM}(h\rightarrow b\bar{b})}
\end{eqnarray}
Here we do not use the LHC results of $h\to
\tau^{+}\tau^{-}$ due to the large
uncertainty. We also note that an excesses in $Vbb$ channel was
also observed by the CMS collaboration, but its value was below the
expectations for a SM Higgs boson at 125 GeV. So we only consider
the measurement of $Vbb$ at Tevatron \cite{tev-vbb,tev-higgs}, which is
$R_{Vbb}^{Tev}=1.97+0.74-0.68$. For the observed Higgs boson with a
mass of $125-126$ GeV, the best fits to the signal rates
are given by
\cite{NovAtlas,DecAtlas,NovCMS}
\small
    \begin{eqnarray}
   && R_{ZZ^*}^{ATLAS}=1.01+0.45-0.40,~R_{ZZ^*}^{CMS}=0.81+0.35-0.28,~R_{ZZ^*}^{\rm combined}=0.88\pm0.25;\nonumber \\
   && R_{WW^*}^{ATLAS}=1.35+0.57-0.52,~R_{WW^*}^{CMS}=0.70+0.25-0.23,~R_{WW^*}^{\rm combined}=0.80\pm0.22; \nonumber \\
   && R_{\gamma\gamma}^{ATLAS}=1.77+0.41-0.38,~R_{\gamma\gamma}^{CMS}=1.56+0.46-0.42,~R_{\gamma\gamma}^{\rm combined}=1.68\pm0.29.~~~~~~~
    \end{eqnarray}
\normalsize
In our study, we require the theoretical predictions
in our model to agree with the combined data at $1\sigma$ level.
We note that very recently, the CMS collaboration has measured $h\to
Z\gamma$ and set an upper limit on the production rate \cite{cms-zr}. We
also include this bound in our study.
\end{itemize}

In the numerical calculations, we take the SM parameters as \cite{pdg}
\begin{eqnarray}
m_t=175{\rm ~GeV},~m_{Z}=91.19 {\rm
~GeV},~\sin^{2}\theta_W=0.2228,~\alpha=1/128.
\end{eqnarray}
We use the parton distribution function CTEQ6m \cite{cteq}
with renormalization
scale and factorization scale $\mu_R = \mu_F = m_t$ for $t\bar{t}$
production. We scan the parameters in the following ranges
\begin{eqnarray}
&& 100{\rm ~GeV}<m_{\phi}<500 {\rm ~GeV}, ~0.5<f_{ut}<1.2 \nonumber \\
&& -10<g_{h\phi\phi}<10,~-\frac{\pi}{2}<\alpha<\frac{\pi}{2},~5<\tan\beta<50.
\end{eqnarray}
Here the values of upper and lower bounds of $m_{\phi}$ and $f_{ut}$
are set to satisfy the latest search for top+jet resonance at the
LHC \cite{atlas-tj}. For the coupling $g_{h\phi\phi}$,
we find it can be as large as ten without conflicting with
the unitarity constraints \cite{unitarity}.

\begin{figure}[htbp]
\includegraphics[width=16cm,height=4in]{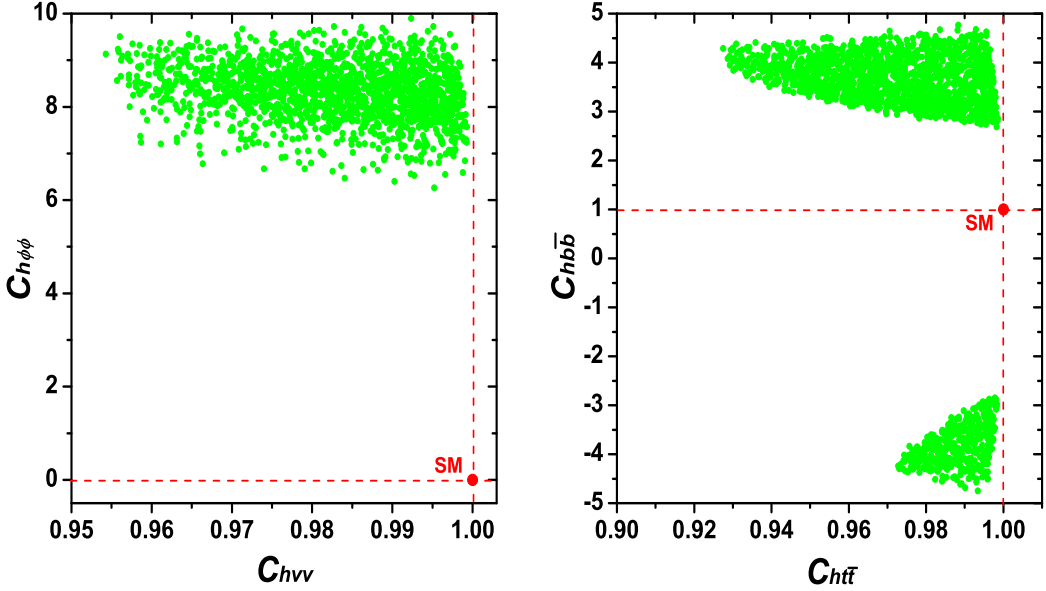}
\vspace{-1.5cm} \caption{The scatter plots of the parameter space
allowed by the experimental constraints, projected on the planes of
$g_{h\phi\phi}$ versus $c_{hVV}$ and $c_{hb\bar{b}}$ versus
$c_{ht\bar{t}}$. Here the couplings $c_{hVV}$, $c_{hb\bar{b}}$ and
$c_{ht\bar{t}}$ are normalized to the SM values.}
 \label{fig1}
\end{figure}
In Fig.1, we project the samples that survive the experimental
constraints (i)-(iv) on the planes of $g_{h\phi\phi}$ versus $c_{hVV}$
and $c_{hb\bar{b}}$ versus $c_{ht\bar{t}}$.
We see that the couplings $c_{hVV}$ and $c_{ht\bar{t}}$ in our model
are very close to their SM values, but $c_{hb\bar{b}}$ can be
much larger than the SM value. This result is very different from the usual
type-II THDM \cite{THDM}, where the diphoton rates are
usually enhanced by reducing the $hb\bar{b}$ coupling.
It can also be seen that the samples with a positive $c_{hb\bar{b}}$
can allow for a smaller $c_{ht\bar{t}}$. This is because the
negative mixing angle $\alpha$ can make the coupling of $c_{hVV}$
larger and a small $c_{ht\bar{t}}$ can suppress
the production of $gg\to h$ to make $h\to VV^{*}$
consistent with the LHC data.

\begin{figure}[htbp]
\includegraphics[width=6.5in,height=6in]{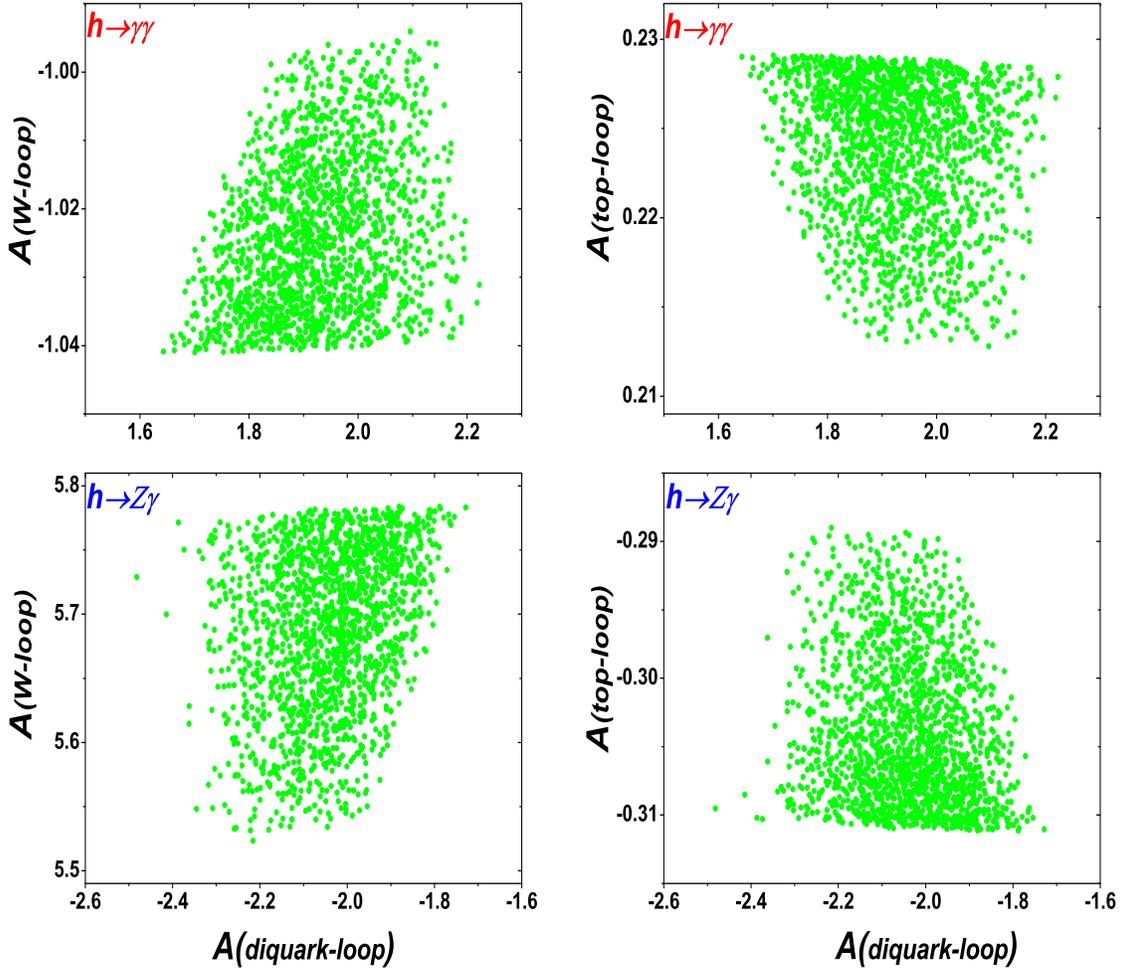}
\vspace{-2.3cm} \caption{Same as Fig.1, but showing
the contribution of $W$-loop,
top-loop and diquark-loop to the amplitude of
$h\to \gamma\gamma$ and $h\to Z\gamma$.}
\label{fig2}
\end{figure}

From Fig.1 we also see that the coupling $g_{h\phi\phi}$ must be larger than 6.3
to meet the diphoton rates observed at the LHC.
This can be understood from Fig.2, which shows
the contribution of $W$-loop,
top-loop and diquark-loop to the amplitude of
$h\to \gamma\gamma$ and $h\to Z\gamma$.
We can see that the
diquark-loop contribution to the amplitude of $h \to \gamma\gamma$
is constructive to the top-loop but destructive to the $W$-loop.
In order to enhance the diphoton rate, the diquark-loop should
be dominant over the $W$-loop. We also note that with
the increase of the diquark-loop, the $W$-loop
and top-loop will be reduced by the couplings of $c_{hVV}$ and
$c_{ht\bar{t}}$ so as not to cause excessive production of $gg \to h
\to VV$ at the LHC.
For $h\to Z\gamma$, the diquark-loop has the same sign as the top-loop.
But different from $h\to \gamma\gamma$, the $W$-loop contribution
to the amplitude of  $h\to Z\gamma$ is always dominant even for a
large $g_{h\phi\phi}$.

\begin{figure}[htb]
\includegraphics[width=16cm]{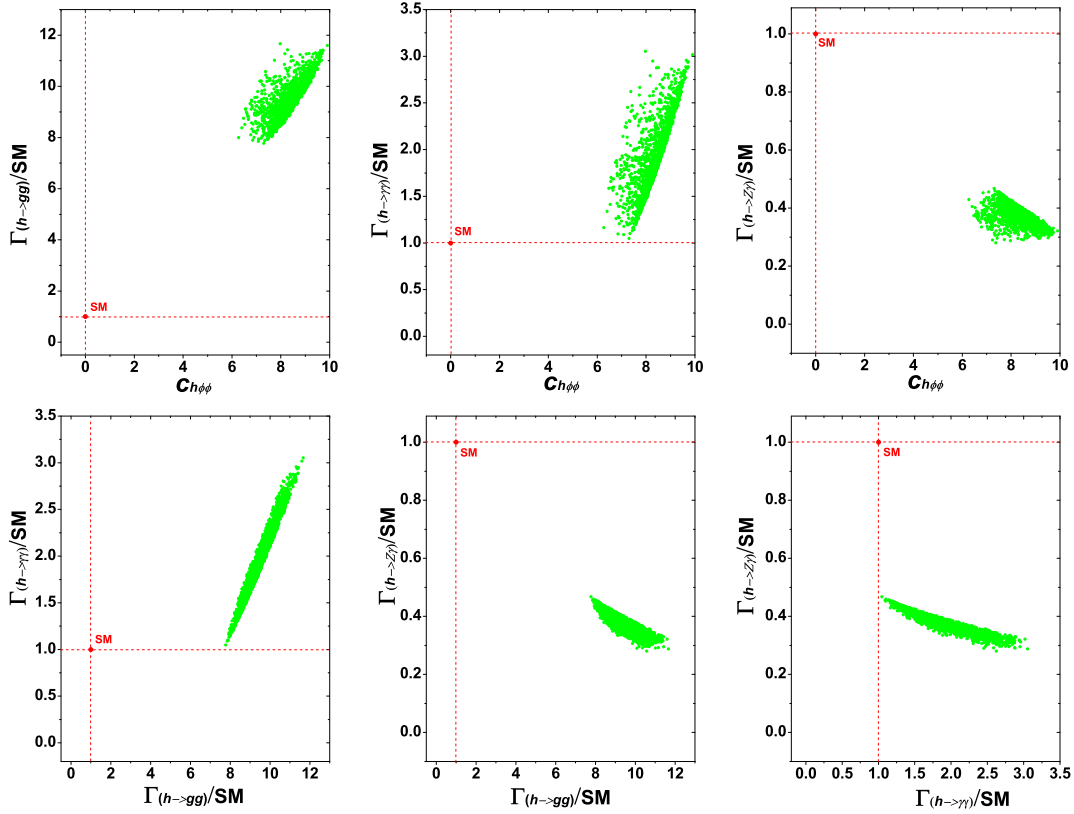}
\vspace{-1cm} \caption{Same as Fig.1, but showing the widths of
$\Gamma(h\to \gamma\gamma)$, $\Gamma(h\to gg)$ and
$\Gamma_{Z\gamma}$ normalized to the SM values. } \label{fig3}
\end{figure}
In Fig.3 we show the Higgs decay widths and their correlations.
We can see that as the coupling $g_{h\phi\phi}$ increases,
$\Gamma(h\to gg)$ and $\Gamma(h\to\gamma\gamma)$ become large,
which can maximally exceed the SM values by a factor of
12 and 3 respectively.
But $\Gamma(h\to Z\gamma)$ decreases and drops to $30\%$ of the SM
value when $g_{h\phi\phi}$ goes up.
These features lead to the different correlation
behaviors, as shown in the below panel of Fig.3.
The reasons for these features are two folds.
One is that  for a large $g_{h\phi\phi}$ the diquark-loop is dominant
in the decays $h\to\gamma\gamma$ and $h\to gg$.
Meanwhile, the sign of
the diquark-loop is same as the top-loop so that it can greatly
increase the width of $h\to gg$ and $h\to\gamma\gamma$
when $g_{h\phi\phi}$ gets large.
However, we should mention that the effect of a large
$\Gamma(h\to\gamma\gamma)$ on increasing the branch ratio
$Br(h\to\gamma\gamma)$ is limited, because the main partial
width of $\Gamma(h\to b\bar{b})$ is also enhanced by a
large $\tan\beta$.
Therefore, in our model the feasible way to increase the diphoton
rates is to enhance the cross section of $gg\to h$.
The other reason is that unlike the case in
$h\to \gamma\gamma$, the diquark-loop in $h\to Z\gamma$ can only
cancel a small part of the contribution of $W$-loop.
Thus, as $g_{h\phi\phi}$ becomes large, the neat combined
contribution to  $h\to Z\gamma$ gets smaller in magnitude.

Finally, in Fig.4 we show the results of $A^t_{FB}$ at the Tevatron
correlated with the LHC Higgs diphoton and $Z$-photon signal rates.
The recent top quark measurement at the Tevatron gives
$A^{t}_{FB} = 15.0 \pm 5.5 \%$ \cite{tev-afb}, which is larger than
the SM prediction 0.056(7) \cite{afb-sm}.
Note that the diquark contributes to the
$t\bar{t}$ production through $u$-channel and will distort
the $t\bar{t}$ invariant mass distribution.
The measurement of this $t\bar{t}$
invariant mass distribution was performed by
CDF \cite{tev-tt-inv} and ATLAS collaborations \cite{Aad:2012hg}.
We require the new physics contribution in each bin to lie within the
$2\sigma$ range of the experimental values.
However, because the shape of such a distribution in the high
energy tail is sensitive to the cut efficiency of event selection
and also sensitive to QCD corrections, we also show the results
without considering the constraints from this distribition.

From Fig.4 we see that without the constraints of $m_{t\bar{t}}$, the
prediction of $A^{t}_{FB}$ at the Tevatron and the Higgs diphoton
rate at the LHC can simultaneously lie in the $1\sigma$ ranges of the
experimental values. We also note that the $Z\gamma$ rate of the Higgs
is suppressed, below the half value of the SM prediction.
Therefore, the measurement of $Z\gamma$ rate will be
useful for the test of our model.

\begin{figure}[htbp]
\includegraphics[width=6in,height=4in]{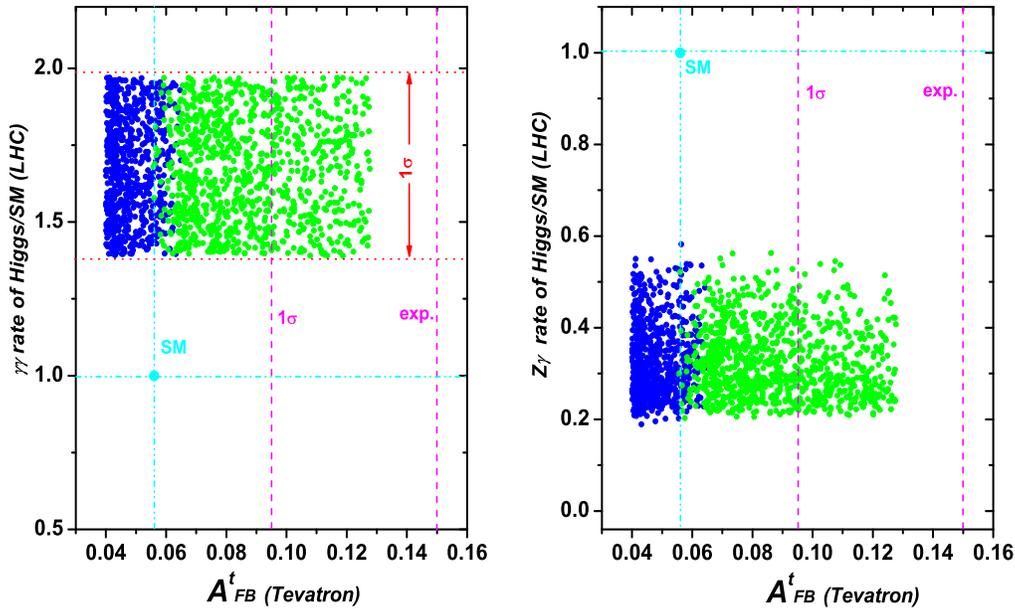}
\vspace{-1cm} \caption{Same as Fig.1, but showing the results of
$A^t_{FB}$ at the Tevatron correlated with the LHC Higgs diphoton
and $Z$-photon signal rates. The bullets (blue) are the samples with
the $m_{t\bar{t}}$ constraints; while the times (green) are the
samples without the $m_{t\bar{t}}$ constraints.} \label{fig4}
\end{figure}

\section{Conclusion}

In this paper we studied the Higgs boson decays to $\gamma\gamma$ and
$Z\gamma$ and the top quark forward-backward asymmetry in a
two-Higgs-doublet model with a color-triplet scalar. We found
that under the current experimental constraints from the Higgs data
and the top quark measurements, such a
model can explain at $1\sigma$ level the anomaly of the top quark
forward-backward asymmetry observed by the Tevatron
and the diphoton enhancement of the Higgs boson observed by
the LHC. We also checked the correlation between $h\to
\gamma\gamma$, $h\to Z\gamma$ and $A_{FB}$ and found that the decay
width of $h\to Z\gamma$ will be highly suppressed due to the
cancelation between diquark-loop and $W$-loop. Therefore, the
future measurement of $h\to Z\gamma$ at the LHC will help
to test our model.

\section*{Acknowledgement}
We thank Prof. Junjie Cao for discussions. C. Han was supported by
a visitor program of Henan Normal University,
during which this work was finished.
This work was supported in part
by the National Natural Science Foundation of China under
grant Nos. 11275245, 10821504 and 11135003, by the Project of Knowledge
Innovation Program (PKIP) of Chinese Academy of Sciences under grant
No. KJCX2.YW.W10. and by the Startup Foundation for Doctors of Henan
Normal University under contract No.11112.

\end{document}